\documentclass[twocolumn,showpacs,pre,aps,preprintnumbers,floatfix,amssymb,showkeys]{revtex4}
\usepackage{graphicx,amsmath,amsfonts}
\renewcommand{\Re}{\mathop{\rm Re}\nolimits}
\renewcommand{\Im}{\mathop{\rm Im}\nolimits}

\newcommand{\Div}{\mathop{\rm div}\nolimits}

\newcommand{\ex}{\mathop{\rm e}\nolimits}
\newcommand{\exi}[1]{\mathop{\rm e}\nolimits^{{\mathrm i}#1}}
\newcommand{\cj}{\mathop{\rm c.c.}\nolimits}

\newcommand{\irm}{\mathrm i}
\begin{document}
\preprint{APS/EU?}
\title{Amplitude equations for 3D double-diffusive convection interacted with a horizontal
vortex}
\author{S.B. Kozitskiy}
\email{skozi@poi.dvo.ru}
\affiliation{Laboratory of Geophysical Hydrodynamics, V.I. Il'ichev
Pacific Oceanological Institute of the Russian Academy of Sciences,
690041, Baltiyskaya 43, Vladivostok, Russia}
\date{\today}
\begin{abstract}
Three dimensional roll-type double-diffusive convection in a
horizontally infinite layer of an uncompressible liquid is
considered in the neighborhood of Hopf bifurcation points. A system
of amplitude equations for the variations of convective rolls
amplitude is derived by multiple-scaled method. An attention is paid
to an interaction of convection and horizontal vortex. Different
cases of the derived equations are discussed.
\end{abstract}



\maketitle

\section{Introduction}

Physical systems, in which double-diffusion induced convection plays
an essential role, are often found in nature. There are two
components with significantly different diffusion coefficients in
such systems. It can be heat and salt in the sea water, heat and
helium in stellar atmospheres, or two reagents in chemical reactors.
As a result of various spatial distribution of these components in a
gravitational field arises convection, which can have various forms
and lead to a variety of phenomena~\cite{Turner:1974}. Widely known,
for example, salt fingers are, arising in salted and warmed from
above water. It is understandable that the results of
double-diffusive convection, for example, in the ocean, can be
applied to double-diffusive convection in astrophysical systems or
in chemical reactor.

There are a number of works devoted to various theoretical models
of systems with double-diffusive convection.
In 80--90 years the formation of structures in the neighborhood of
Hopf bifurcation points for the horizontally translation-invariant systems
was actively studied in some works.
The development of oscillations in such systems give rise to different
types of waves (eg, standing, running, modulated, chaotic), which
well described by a generalized Ginzburg--Landau equations~\cite{K10,BB}.
The equations of this type must be derived from the basic system of
partial differential equations for the given physical system by asymptotic
methods.
However, a full and well-grounded derivation of amplitude equations for
systems with double-diffusive convection (especially three-dimensional)
is still poorly represented in the literature.

The purpose of this work is the derivation of amplitude equations for
the three-dimensional double-diffusive system in the neighborhood of
Hopf bifurcation points for the case of roll-type convection. This extends the idea
of previous work~\cite{K10,K11}, where a two-dimensional and three-dimensional
convection in a square-cells was investigated by alike methods.

\section{Formulation of the Problem and Basic Equations}

Consider 3D double-diffusive convection in a liquid layer of
a width $h$, confined by two plane horizontal boundaries.
The liquid layer is heated and salted from below.
The governing equations in this case are hydrodynamical equations for
a liquid mixture in the gravitational field~\cite{Lan}:

\begin{equation*}\label{meq}
\begin{split}
& \partial_t{\bf v} + ({\bf v}\nabla){\bf v}=
  -\rho^{-1}\nabla p + \nu \Delta{\bf v}+{\bf g}\,,\\
& \partial_t T + ({\bf v}\nabla) T = \chi\Delta T\,,\\
& \partial_t S + ({\bf v}\nabla) S = D\Delta S\,,\\
& \Div{\bf v} = 0\,.
\end{split}
\end{equation*}

Where ${\bf v}(t,x,y,z)$ is the velocity field of liquid, $T(t,x,y,z)$
is the temperature, $S(t,x,y,z)$ is the salt concentration, $p(t,x,y,z)$
is the pressure, $\rho(t,x,y,z)$ is the density of liquid, ${\bf g}$
is the acceleration of gravity, $\nu$ is the kinematic viscosity of fluid,
$\chi$ is the thermal diffusivity of the liquid, $D$ is the salt diffusivity.
Cartesian frame with the horizontal $x$-axis and $y$-axis is used, while
the $z$-axis is directed upward and $t$ is the time variable.

Distributed sources of heat and salt are absent.
On the upper and lower boundaries of the layer
the constant values of temperature and salinity are supported,
higher at the lower boundary.
The governing equations are transformed into dimensionless form with the
use of Boussinesq approximation and following units for length, time,
velocity, pressure, temperature and salinity respectively: $h$,
$h^2/\chi$, $\chi/h$, $\rho_0\chi^2/h^2$, $T_{\Delta}$, $S_{\Delta}$, where
$T_{\Delta}$ and $S_{\Delta}$ are temperature and salinity differences across
the layer.

The dimensionless governing equations for momentum and diffusion
of temperature and salt are~\cite{K11}:
\begin{equation}\label{meq3}
\begin{split}
& u_t+(uu_x+vu_y+wu_z)=-p_x+\sigma\Delta u,\\
& v_t+(uv_x+vv_y+wv_z)=-p_y+\sigma\Delta v,\\
& w_t+(uw_x+vw_y+ww_z)=\\
& \qquad\qquad\qquad -p_z+\sigma\Delta w +\sigma R_T\theta-\sigma R_S\xi,\\
& \theta_t+(u\theta_x+v\theta_y+w\theta_z)-w=\Delta\theta,\\
& \xi_t+(u\xi_x+v\xi_y+w\xi_z)-w=\tau\Delta\xi,\\
& u_x + v_y + w_z = 0.
\end{split}
\end{equation}
Where $\sigma=\nu/\chi$ is the Prandtl number
($\sigma\approx7.0$), $\tau=D/\chi$ is the Lewis number ($0<\tau< 1$,
usually $\tau=0.01-0.1$). $R_T =
({{g}{\alpha'}{h^{3}}}/{\chi\nu})T_{\Delta}$ is the temperature
Rayleigh number and $R_S =
({{g}{\gamma'}{h^{3}}}/{\chi\nu})S_{\Delta}$ is the salinity Rayleigh number.

Free-slip boundary conditions are used for the dependent variables
(the horizontal velocity component is undefined):
\begin{equation*} \label{econ}
u_z = v_z = w = \theta = \xi = 0\,\,\mbox{at}\,\, z=0,\,\, 1\,.
\end{equation*}
It is believed that they are suitable to describe the convection in
the inner layers of liquid and do not change significantly the
convective instability occurrence criteria for the investigated class
of systems~\cite{Weiss:1981}.

\section{Derivation of Amplitude Equations -- General Frame of Decomposition}
Consider the equations for double-diffusive convection in the vicinity
of a bifurcation point, the temperature and salinity Rayleigh numbers
for which are designated as $R^{*}_T$ and $R^{*}_S$ respectively.
In this case the Rayleigh numbers can be represented as follows:
\begin{equation*}
R_T=R^{*}_T(1+{\varepsilon}^{2}r_T),
 \qquad
R_S=R^{*}_S(1+{\varepsilon}^{2}r_S).
\end{equation*}
Values of $r_T$ and $r_S$ are of unit order, and the small parameter $\varepsilon$
shows how far from the bifurcation point the system is.
To derive the amplitude equations we use the derivative-expansion method,
which is a variant of multiple-scale method~\cite{Dodd:1982,Nayfeh:1976}.
Introduce the slow variables:
\begin{equation*}
T_1 = \varepsilon t,\quad T_2=\varepsilon^2 t,\quad X_1 =
\varepsilon x,\quad Y_1 = \sqrt{\varepsilon} y.
\end{equation*}
In accordance with the method chosen, we assume that the dependent variables
now depend on $t$, $T_1$, $T_2$, $x$, $y$, $z$, $X_1$, $Y_1$, which are
considered independent. Also replace the derivatives in equations (\ref{meq3})
for the prolonged ones by the rules:
\begin{equation*}  \label{deri}
\partial_t \rightarrow \partial_t + \varepsilon\partial_{T_1}
 + {\varepsilon}^{2}\partial_{T_2}, \quad
\partial_x \rightarrow \partial_x + \varepsilon\partial_{X_1},\quad
\partial_y \rightarrow \sqrt{\varepsilon}\partial_{Y_1}.
\end{equation*}
Then the equations (\ref{meq3}) can be written as:
\begin{equation}\label{meq4}
\begin{split}
& u_t+(uu_x+wu_z)+p_x-\sigma(u_{xx}+u_{zz})= -\sqrt{\varepsilon}vu_{Y_1}\\
& \qquad -\varepsilon [u_{T_1}+uu_{X_1}+p_{X_1}-2\sigma
u_{xX_1}-\sigma u_{Y_1 Y_1}]\\
& \qquad +\varepsilon^2[\sigma u_{X_1 X_1}-u_{T_2}],\\
& v_t+(uv_x+wv_z)-\sigma(v_{xx}+v_{zz})=-\sqrt{\varepsilon}(vv_{Y_1}+p_{Y_1})\\
& \qquad -\varepsilon [v_{T_1}+uv_{X_1}-2\sigma v_{xX_1}-\sigma
v_{Y_1 Y_1}]\\
& \qquad +\varepsilon^2[\sigma v_{X_1 X_1}-v_{T_2}],\\
& w_t+(uw_x+ww_z)+p_z-\sigma (w_{xx}+w_{zz})\\
& \qquad -\sigma R_T\theta+\sigma R_S\xi=-\sqrt{\varepsilon}vw_{Y_1}\\
& \qquad -\varepsilon [w_{T_1}+uw_{X_1}-2\sigma w_{xX_1}-\sigma w_{Y_1 Y_1}]\\
& \qquad+\varepsilon^2[\sigma w_{X_1 X_1}-w_{T_2}+\sigma r_T\theta-\sigma r_S\xi],\\
& \theta_t+(u\theta_x+w\theta_z)-w-(\theta_{xx}+\theta_{zz})=-\sqrt{\varepsilon}v\theta_{Y_1}\\
& \qquad -\varepsilon [\theta_{T_1}+u\theta_{X_1}-2\theta_{xX_1}-
\theta_{Y_1 Y_1}]\\
& \qquad +\varepsilon^2[\theta_{X_1 X_1}-\theta_{T_2}],\\
& \xi_t+(u\xi_x+w\xi_z)-w-\tau(\xi_{xx}+\xi_{zz})=-\sqrt{\varepsilon}v\xi_{Y_1}\\
& \qquad -\varepsilon [\xi_{T_1}+u\xi_{X_1}-2\tau\xi_{xX_1} -
\tau\xi_{Y_1 Y_1} ]\\
& \qquad +\varepsilon^2[\tau\xi_{X_1 X_1}-\xi_{T_2}],\\
& u_x + w_z = -\sqrt{\varepsilon}v_{Y_1}-\varepsilon u_{X_1}.
\end{split}
\end{equation}
We seek solutions of these equations in the form of asymptotic series
in powers of small parameter $\varepsilon$:
\begin{equation}\label{sets}
\begin{split}
&u = \varepsilon u_1+\varepsilon^2 u_2+\varepsilon^3 u_3+\cdots\,,\\
&v = \varepsilon\sqrt{\varepsilon}
v_1+\varepsilon^2\sqrt{\varepsilon} v_2
    +\varepsilon^3\sqrt{\varepsilon} v_3+\cdots\,,\\
&w = \varepsilon w_1+\varepsilon^2 w_2+\varepsilon^3 w_3+\cdots\,,\\
&p = \varepsilon p_1+\varepsilon^2 p_2+\varepsilon^3 p_3+\cdots\,,\\
&\theta =
\varepsilon\theta_1+\varepsilon^2\theta_2+\varepsilon^3\theta_3+\cdots\,,\\
&\xi =
\varepsilon\xi_1+\varepsilon^2\xi_2+\varepsilon^3\xi_3+\cdots\,.
\end{split}
\end{equation}
After their substitution in (\ref{meq4}) and collection the terms at $\varepsilon^n$
we obtain the systems of equations to determine the terms of the series (\ref{sets}).

\section{Terms of the First Order in $\varepsilon$}
At $O(\varepsilon^1)$ we obtain the following system:
\begin{equation}\label{seq1}
\begin{split}
& u_{1t}+p_{1x}-\sigma (u_{1xx}+u_{1zz}) = 0,\\
& v_{1t}+p_{1Y_1}-\sigma (v_{1xx}+v_{1zz}) = 0,\\
& w_{1t}+p_{1z}-\sigma (w_{1xx}+w_{1zz}) -\sigma R_T\theta_1+\sigma R_S\xi_1=0,\\
& \theta_{1t}-w_1-(\theta_{1xx}+\theta_{1zz}) = 0,\\
& \xi_{1t}-w_1-\tau(\xi_{1xx}+\xi_{1zz}) = 0,\\
& u_{1x} + w_{1z} = 0.
\end{split}
\end{equation}
Solutions in the form of normal modes (roll-type cells) are:
\begin{equation}\label{nmode1a}
\begin{split}
& w_1= A\exi{kx}\ex^{\lambda t}\sin \pi z + \cj,\\
& \theta_1 = \bar{\theta}_{1a}\exi{kx}\ex^{\lambda t}\sin \pi z + \cj,\\
& \xi_1 = \bar{\xi}_{1a}\exi{kx}\ex^{\lambda t}\sin \pi z + \cj,\\
& p_1= \bar{p}_{1a}\exi{kx}\ex^{\lambda t}\cos \pi z +\hat{p}_1 + \cj,\\
& u_1= \bar{u}_{1a}\exi{kx}\ex^{\lambda t}\cos \pi z +\hat{u}_1 + \cj,\\
& v_1= \bar{v}_{1a}\exi{kx}\ex^{\lambda t}\cos \pi z +\hat{v}_1 +
\cj.
\end{split}
\end{equation}
Terms with the bars and hats depend only on the slow variables $X_1,Y_1,T_1,T_2$.
Substituting the expressions (\ref{nmode1a}) in the equation (\ref{seq1})
gives relations for the amplitudes. Hereinafter $\varkappa^2=k^2+\pi^2$
is the full wave number.

Parameters $\lambda, k, R_T, R_S$ are related by:
\begin{equation*}\label{disa}
\begin{split}
&(\lambda+\sigma\varkappa^2)(\lambda+\varkappa^2)(\lambda+\tau\varkappa^2)\\
& \qquad + \sigma
(k^2/\varkappa^2)[R_S(\lambda+\varkappa^2)-R_T(\lambda+\tau\varkappa^2)]=0\,.
\end{split}
\end{equation*}
\begin{figure}[tbh]
\begin{center}
\begin{picture}(215,160)
\put(15,10){\vector(0,1){150}} \put(15,10){\vector(1,0){200}}
\put(40,10){\line(1,1){50}} \put(90,60){\line(1,2){50}}
\put(90,60){\circle*{3}} \put(95,55){$C$} \put(0,155){$R_S$}
\put(210,0){$R_T$} \put(50,105){I} \put(165,60){II} \put(55,40){{\it
1}} \put(100,105){{\it 2}}
\end{picture}
\end{center}
\caption{Plane of the Rayleigh numbers $R_T - R_S$. On the line 1 takes place Taylor bifurcation,
on the line 2 takes place Hopf bifurcation.}
\end{figure}
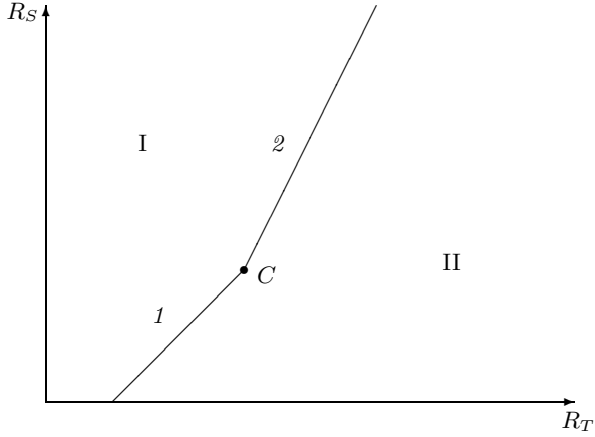
This equation has three roots, two of which can be complex conjugates.
In the case of Hopf bifurcation these two roots acquire positive real part
at some $R_T^*$ ($\omega$ is a frequency of convective waves):
\begin{equation*}
\begin{split}
& R_T^* = \frac{\sigma+\tau}{1+\sigma}R_S^* +
\frac{\varkappa^6}{\sigma k^2}(1+\tau)(\tau+\sigma)\,,\\
& \qquad \omega^2 = \frac{1-\tau}{1+\sigma}\sigma
R_S^*\frac{k^2}{\varkappa^2}-\tau^2\varkappa^4.
\end{split}
\end{equation*}

\section{The Equations at $\varepsilon^2$ and $\varepsilon^3$}
The obtained systems can be written in general form as:
\begin{equation}
\hat{L}\varphi_i = Q_i. \nonumber
\end{equation}
Here $\hat{L}$ is linear differential operator such that $\hat{L}\varphi_1=0$ corresponds
to the system (\ref{seq1}), where $\varphi_i= (u_i,v_i,w_i,\theta_i,\xi_i)$. Functions $Q_i$ include
terms, resonating with the left side of the equations, namely: $Q_i=Q_i^{(1)}+Q_i^{(2)}+Q_i^{(3)}$.
Here $Q_i^{(1)}$ and $Q_i^{(2)}$ generate secular terms of the two types in solutions,
and $Q_i^{(3)}$ does not generate secular terms of any kind.
The condition of absence of secular terms of the first type consists in the requirement
of orthogonality of functions $Q_i^{(1)}$ and the solution of the adjoint homogeneous equation
$\hat{L}^{\star}\varphi_i^{\star} = 0$ \cite{Dodd:1982, Nayfeh:1976},
which usually takes the form of an amplitude equation.
Terms $Q_i^{(2)}$ are constants with respect to the fast variables, and
to exclude the violation of the regularity of the expansions (\ref{nmode1a}),
they should be equated to zero~\cite{BB}. These conditions also have the form of amplitude equations.

As a result, the condition of absence of secular terms of the first type in solution
of equations at $\varepsilon^2$ is as follows ($\alpha_0$ is defined in (\ref{seq5a7}) below):
\begin{equation*}
A_{T_1}+ \alpha_0(2\irm kA_{X_1}+A_{Y_1Y_1})+\irm kA\Omega_{Y_1}=0\,.
\end{equation*}
To exclude secular terms of the second type, we introduce the horizontal stream function
$\Omega$ such that:
\begin{equation*}
\hat u_1 = \Omega_{Y_1},\qquad \hat v_1 = -\Omega_{X_1}\,, \qquad
\Omega_{T_1} - \sigma\Omega_{Y_1Y_1}=0.
\end{equation*}

\begin{figure}[tbh]
\centering
\includegraphics[width=0.34\textwidth,angle=-90]{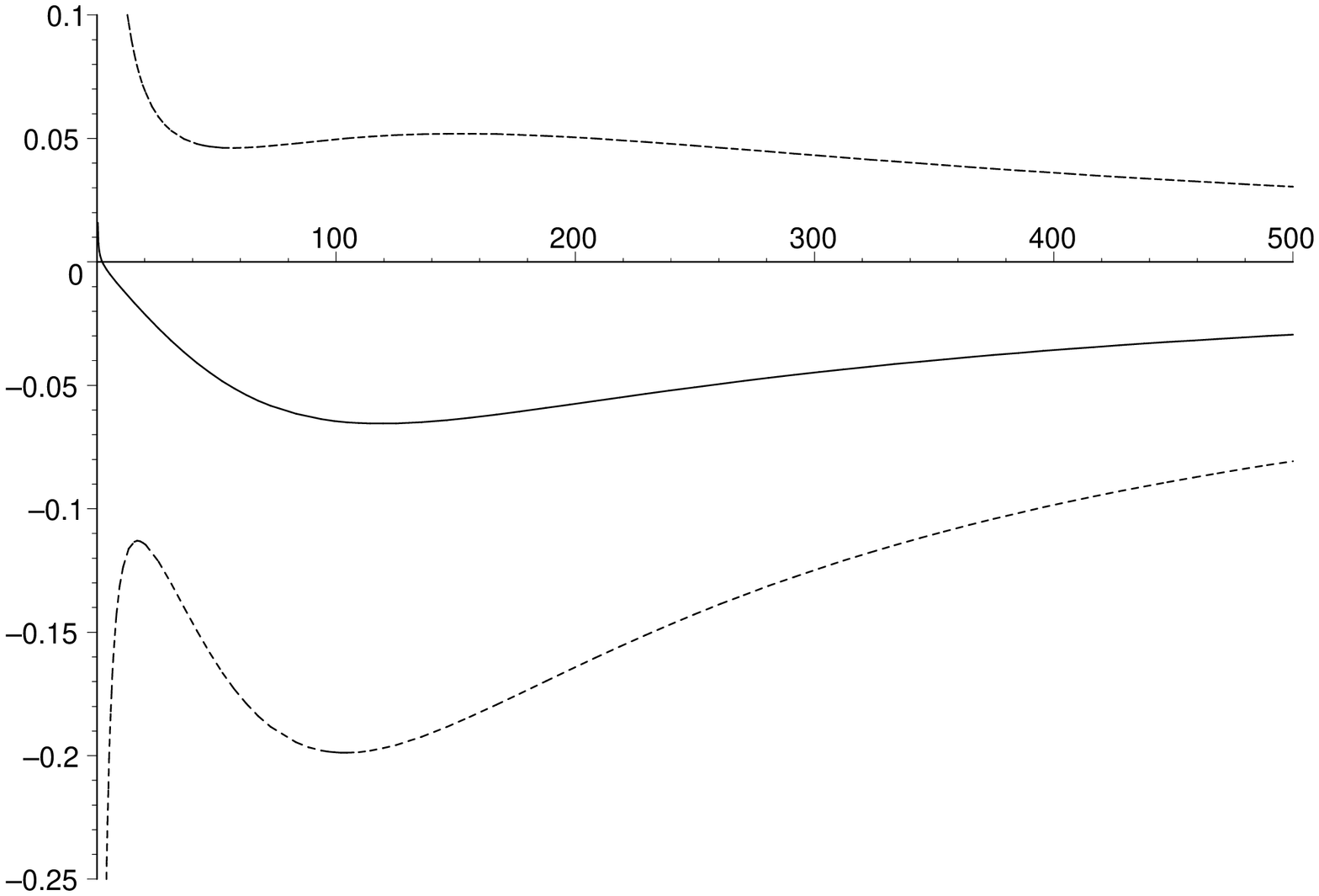}
\includegraphics[width=0.34\textwidth,angle=-90]{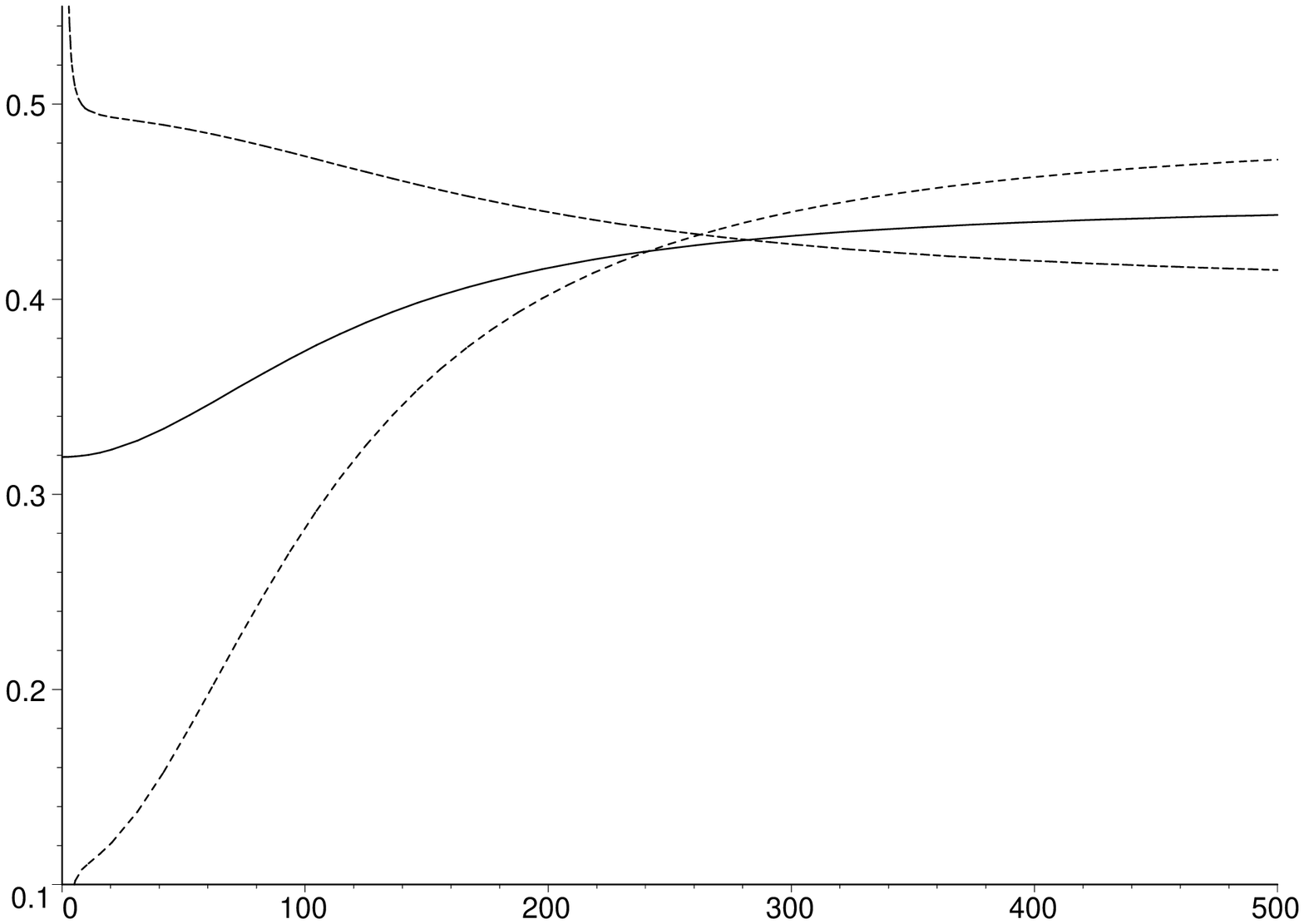}
\caption{\label{fig4} $\Re(\alpha_{3}(\omega))$ (left) and $\Im(\alpha_{3}(\omega))$ (right)
with different $k$: $k=1.75$ (dots), $k=\pi/\sqrt{2}$ (solid), $k=3$ (dash). At $\sigma=7$ and $\tau=0.02$.}
\end{figure}

\begin{figure}[tbh]
\centering
\includegraphics[width=0.34\textwidth,angle=-90]{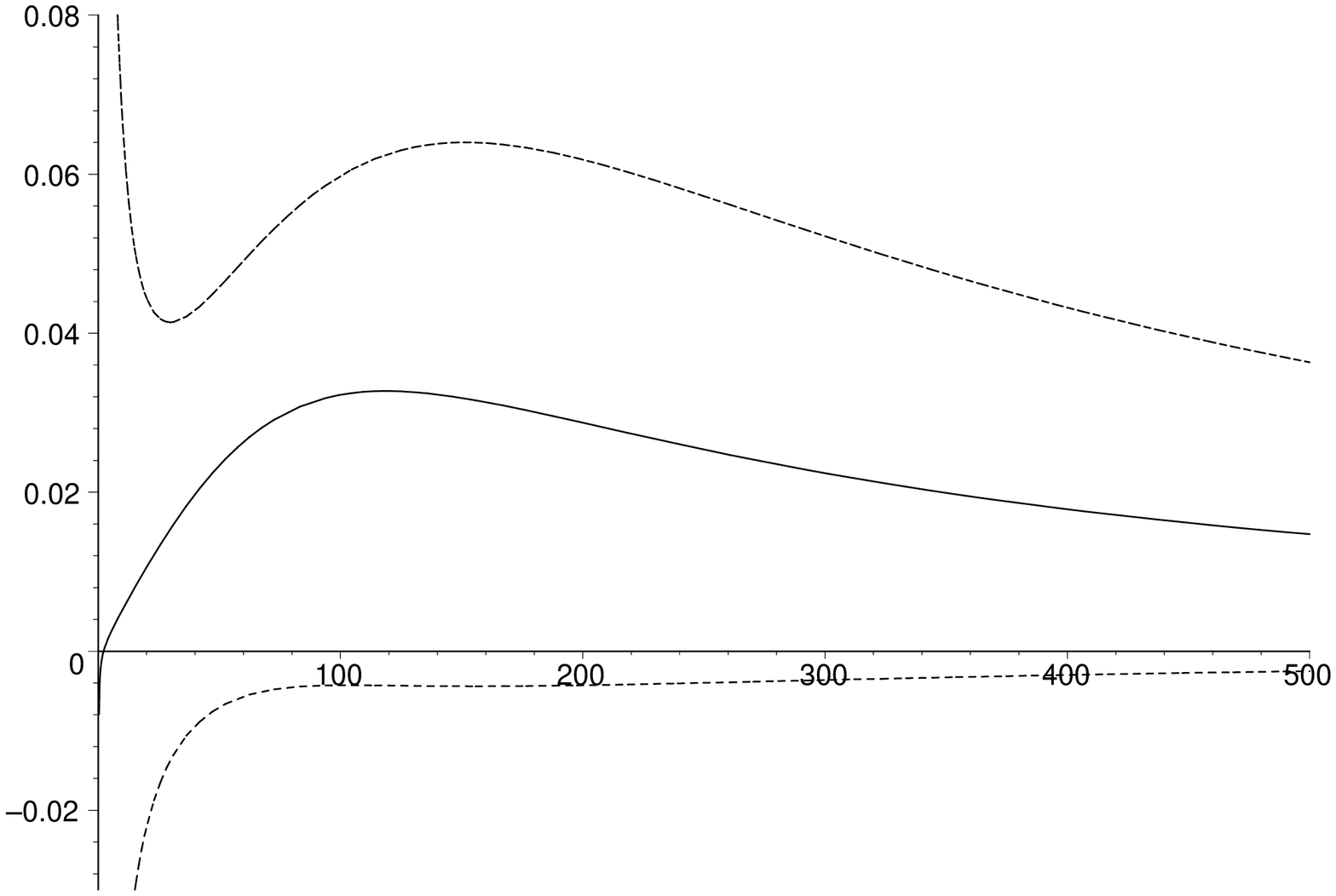}
\includegraphics[width=0.34\textwidth,angle=-90]{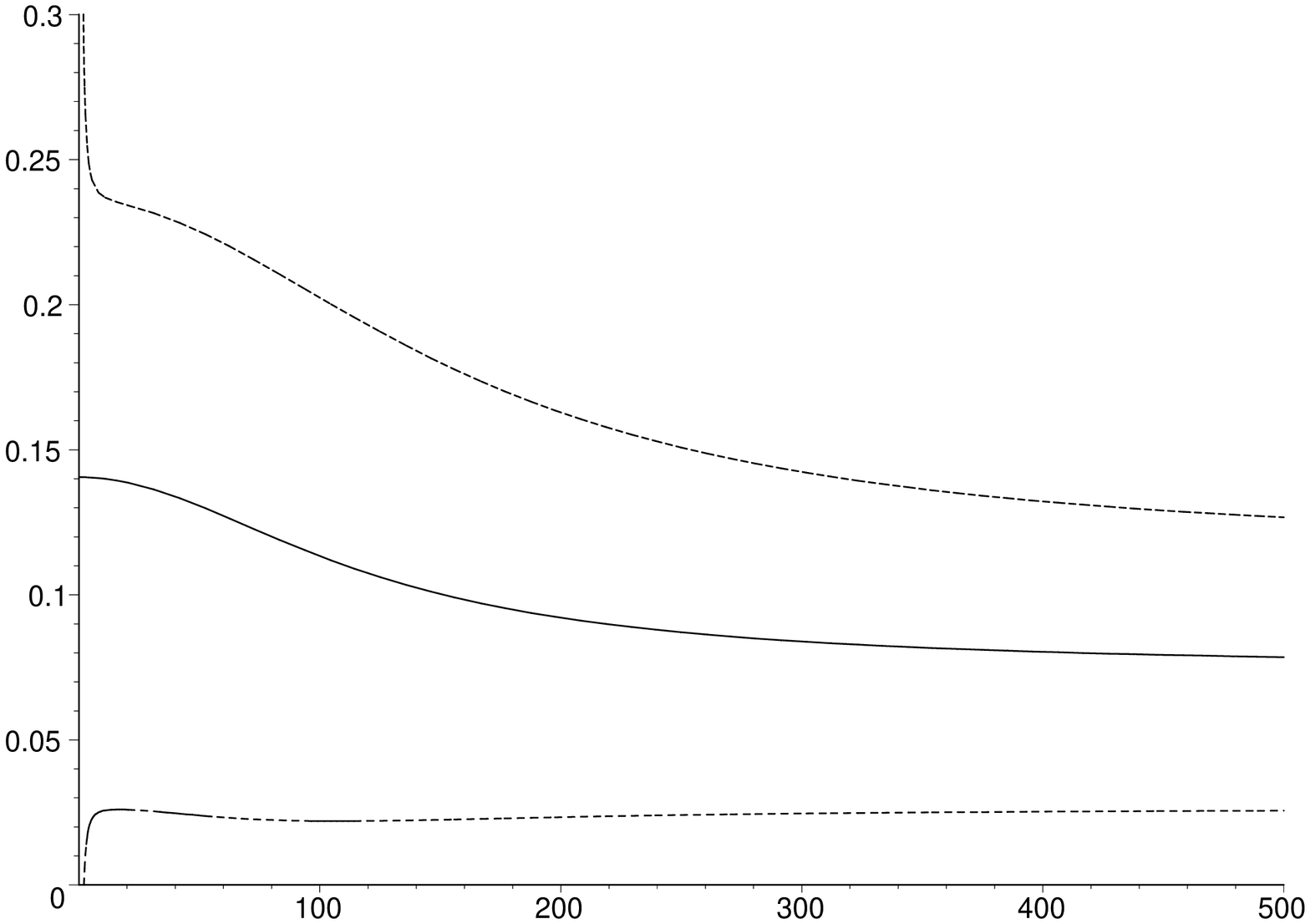}
\caption{\label{fig6} $\Re(\alpha_{4}(\omega))$ (left) and $\Im(\alpha_{4}(\omega))$ (right)
with different $k$: $k=1.75$ (dots), $k=\pi/\sqrt{2}$ (solid), $k=3$ (dash). At $\sigma=7$ and $\tau=0.02$.}
\end{figure}

\section{$A\Omega$-system of Amplitude Equations}
We write the resulting amplitude equations for the system at $\varepsilon^3$:
\begin{equation}\label{seq5a6}
\begin{cases}
A_{T_2}= r A+\alpha_1\left(\partial_{X_1}+\displaystyle{\frac{1}{2\irm k}}\partial_{Y_1}^2\right)^{\!\! 2}\!\! A-\alpha_0 A_{X_1 X_1}\\
\qquad\qquad\qquad + \alpha_2 A|A|^2 + J(\Omega,A) + \hat{F}(\Omega,A)\,, \medskip \\
(\Omega_{T_2}-\sigma\Omega_{X_1 X_1})_{X_1} = J(\Omega,\Omega_{X_1})+\hat{G}(A)\,.\\
\end{cases}
\end{equation}
Here the Jacobian $J(\Omega,f)=(\Omega_{X_1}f_{Y_1}-\Omega_{Y_1}f_{X_1})$ is introduced.
Also operators $\hat{F}(\Omega,A)$ and $\hat{G}(A)$ are defined as follows:
\begin{equation*}
\begin{split}
& \hat{F}(\Omega,A) = \alpha_3(\irm kA\Omega_{X_1Y_1}+A_{Y_1}\Omega_{Y_1Y_1})+\alpha_4 A\Omega_{Y_1Y_1Y_1},\\
%& \qquad +(\Omega_{X_1}A_{Y_1}-\Omega_{Y_1}A_{X_1})\,,\\
& \hat{G}(A) = \frac{\pi^2}{k^4}(|A_{Y_1}|^2)_{Y_1}+\frac{\pi^2}{k^4}\Re(\irm k(AA_{Y_1}^{*})_{X_1})\,.
\end{split}
\end{equation*}
The coefficients of these equations are given by the expressions:
\begin{equation}\label{seq5a7}
\begin{split}
& r = \left(\frac{\sigma k^2}{2\irm\omega\varkappa^2}\right)
\frac{(\irm\omega+\tau\varkappa^2)r_T-(\irm\omega+\varkappa^2)r_S}
{\irm\omega+(1+\tau+\sigma)\varkappa^2} \,,\\
& \alpha_0 = \frac{\irm\omega}{\varkappa^2}
\left[1+\left(\frac{\pi^2}{2k^2}-1\right)\left(1-\frac{\varkappa^4}{\omega^2}\beta_1\right)\right]=\frac{\irm\omega}{\varkappa^2}+\beta ,\\
& \alpha_1 = \left(\frac{\pi^2}{k^2}-1\right)\left(\frac{2\irm\omega}{\varkappa^2}+\frac{2\varkappa^2}{\irm\omega}\beta_1\right)\\
& \qquad -\frac{8k^2}{\varkappa^2}\beta\left[1+\left(\frac{\pi^2}{2k^2}-1\right)\beta_2\right]+\frac{4k^2}{\varkappa^2}\beta^2\beta_3 \,,\\
& \alpha_2 = \frac{\varkappa^2}{4\irm\omega}\,,\quad
 \alpha_3 = \frac{2\irm k}{\varkappa^2}\left[1+\left(\frac{\pi^2}{k^2}-1\right)\beta_2-\beta\beta_3\right] \,,\\
& \alpha_4 = \frac{\irm k}{\varkappa^2}\left(1-\beta_2-\beta\beta_3\right)\,.
\end{split}
\end{equation}
Here for the convenience and compactness of the formulas we introduced functions:
\begin{equation*}\label{seq5a8}
\begin{split}
& \beta = \frac{\irm\omega}{\varkappa^2}\left(\frac{\pi^2}{2k^2}-1\right)\left(1-\frac{\varkappa^4}{\omega^2}\cdot
\frac{(\tau+\sigma+\tau\sigma)\irm\omega+\tau\sigma\varkappa^2}{\irm\omega+(1+\tau+\sigma)\varkappa^2}\right),\\
& \beta_1 = \frac{(\tau+\sigma+\tau\sigma)\irm\omega+\tau\sigma\varkappa^2}
{\irm\omega+(1+\tau+\sigma)\varkappa^2}\,, \\
& \beta_2 = \frac{(\irm\omega+\varkappa^2)(\irm\omega+\tau\varkappa^2)}
{2\irm\omega(\irm\omega+(1+\tau+\sigma)\varkappa^2)}\,,\\
& \beta_3 = -\frac{\varkappa^2(\omega^2+\tau\varkappa^4)}{2\irm\omega(\irm\omega+\varkappa^2)(\irm\omega+\tau\varkappa^2)}\\
& \qquad\qquad +\frac{\varkappa^4((1+\tau+2\sigma)\irm\omega+(1+\tau^2+\tau\sigma+\sigma)\varkappa^2)}{2(\irm\omega+\varkappa^2)
(\irm\omega+\tau\varkappa^2)(\irm\omega+(1+\tau+\sigma)\varkappa^2)}\,.
\end{split}
\end{equation*}
It is easy to see that for $k=\pi/\sqrt{2}$, which corresponds to the fastest
growing mode for not too high frequencies, the parameter $\beta$ vanishes,
and the above formulas are significantly simplified.
\begin{equation*}\label{seq5a7a}
\begin{split}
& r = \left(\frac{\sigma}{6\irm\omega}\right)\frac{(2\irm\omega+3\tau\pi^2)r_T-(2\irm\omega+3\pi^2)r_S}
{2\irm\omega+3(1+\tau+\sigma)\pi^2}\,, \\
&\alpha_0 = \frac{2\irm\omega}{3\pi^2}\,, \qquad
\alpha_1 = \frac{4\irm\omega}{3\pi^2}+\frac{3\pi^2}{\irm\omega}\beta_1\,,\qquad
\alpha_2 = \frac{3\pi^2}{8\irm\omega}\,, \\
&\alpha_3 = \frac{\irm 2\sqrt{2}}{3\pi}\left(1+\beta_2\right)\,,\qquad
\alpha_4 = \frac{\irm \sqrt{2}}{3\pi}\left(1-\beta_2\right)\,.
\end{split}
\end{equation*}

\section{Complex Ginzburg-Landau (CGL) Equation}
Let us consider some special cases of the system (\ref{seq5a6}) to which it reduces
at various simplifying assumptions. The simplest form of the equation is obtained if
we assume that we are considering:
1). neglect the interaction with the vortex, and
2). consider the dynamics on the single spatial variable, and
3). believe that the wave number corresponds to the first losing stability mode $k=\pi/\sqrt{2}$.
Under these conditions, the system takes the following form:
\begin{equation}\label{seq6a1}
A_{T_2}= r A+\alpha_5 A_{X_1 X_1} +\alpha_2 A|A|^2\,.
\end{equation}
Coefficients of the obtained CGL equation (consistent with~\cite{K10,Bre}) are:
\begin{equation*}\label{seq6a3}
\alpha_5 = \frac{\irm\omega}{\varkappa^2}+\frac{2\varkappa^2}{\irm\omega}\beta_1 =
\frac{\irm\omega}{\varkappa^2}+\frac{2\varkappa^2}{\irm\omega}\cdot
\frac{(\tau+\sigma+\tau\sigma)\irm\omega+\tau\sigma\varkappa^2}
{\irm\omega+(1+\tau+\sigma)\varkappa^2}
\end{equation*}
\begin{figure}[tbh]
\centering
\includegraphics[width=0.34\textwidth,angle=-90]{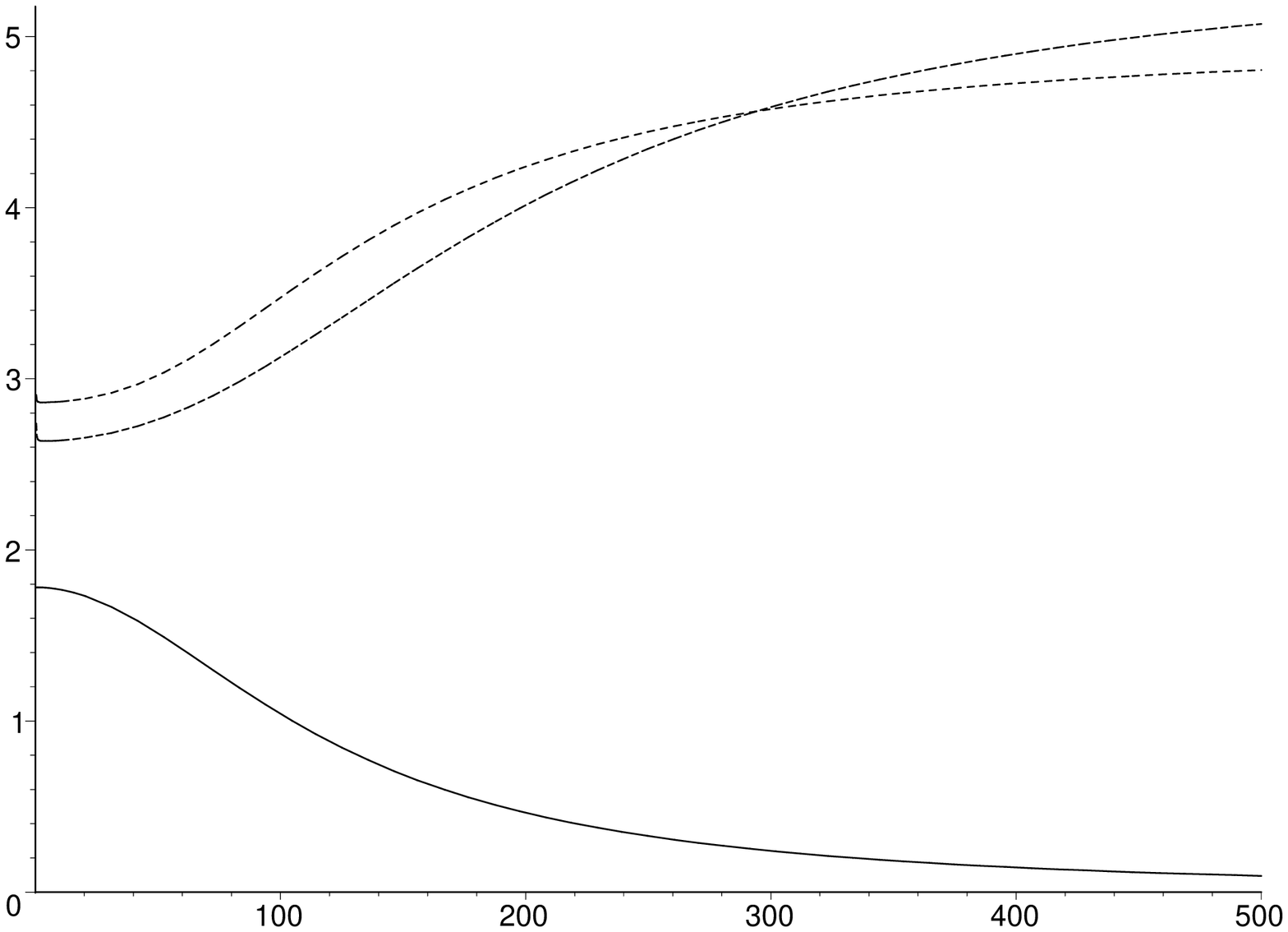}
\includegraphics[width=0.34\textwidth,angle=-90]{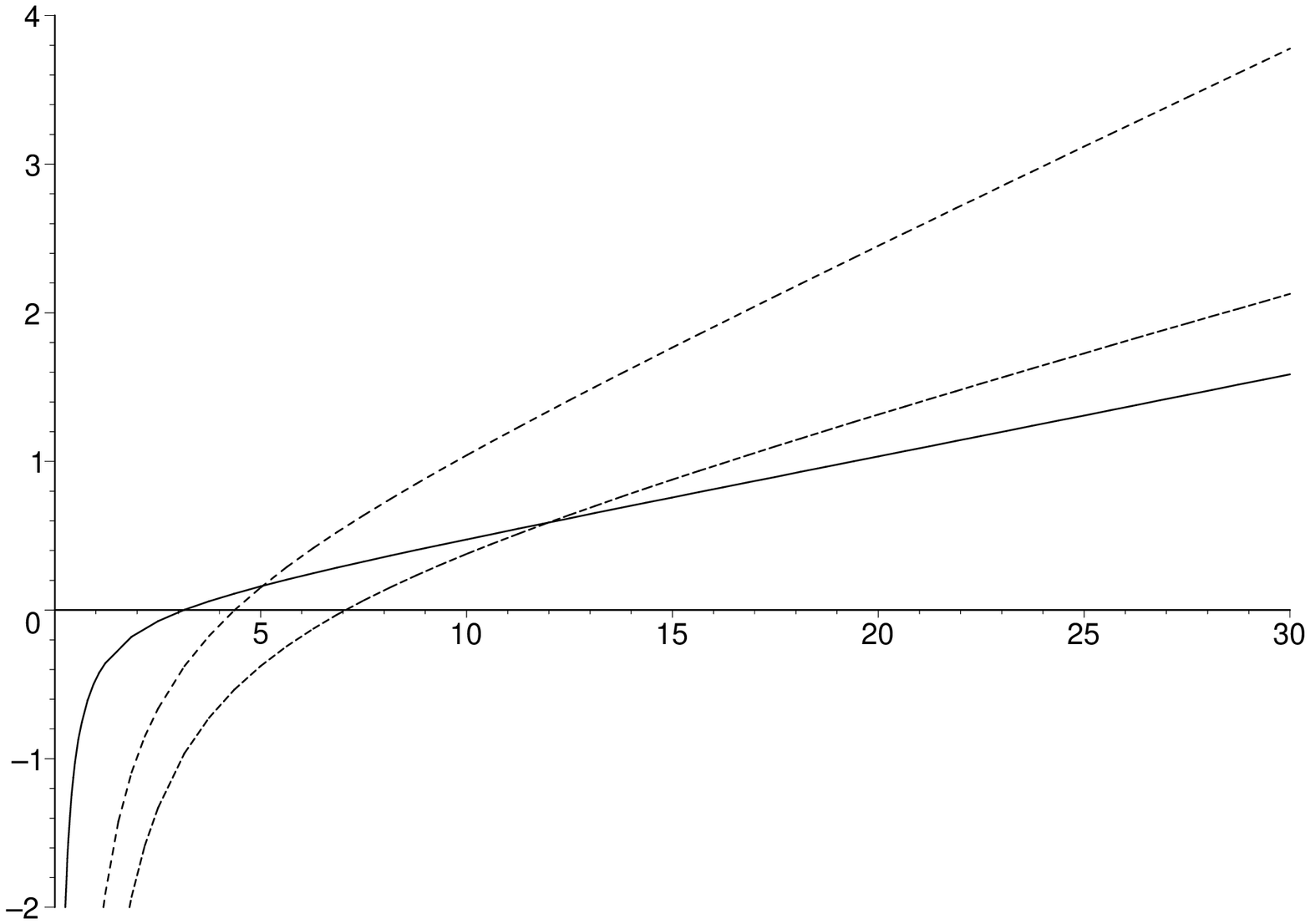}
\caption{\label{fig2} $\Re(\alpha_{5}(\omega))$ (left) and $\Im(\alpha_{5}(\omega))$ (right)
with different $k$: $k=1.75$ (dots), $k=\pi/\sqrt{2}$ (solid), $k=3$ (dash). At
$\sigma=7$ and $\tau=0.02$. }
\end{figure}
In the case when the wave number $k$ is not restricted by any conditions the
coefficient $\alpha_5$ is expressed in general terms as  $\alpha_5= \alpha_1-\alpha_0$
(see fig.~\ref{fig2}).
In the limit of high frequencies $\omega$ the resulting equation reduces to the
nonlinear Schr\"odinger equation and has such solutions as ``dark'' solitons~\cite{K10}.

\section{Conclusion}

\begin{itemize}
\item
The $A\Omega$-system of amplitude equations (\ref{seq5a6}) describing 3D double-diffusive
roll-type convection interacting with horizontal
vorticity field $\Omega$ was derived.
\item
An approach to calculation of amplitude equation coefficients which allows to get
relatively compact formulas such as (\ref{seq5a7}) was developed.
\item
As a special case the complex Ginzburg-Landau equation (\ref{seq6a1}) for the 2D
double-diffusive convection for an arbitrary $k$ (width of the convective cells) was derived.
\end{itemize}
The results can be used to describe the processes of heat and mass transfer,
the formation of vortex structures in the ocean and the atmosphere by convection,
and may also be the basis for constructing more advanced models of this kind.

\begin{acknowledgments}
This work is supported by the grants 09-III-A-07-317 and 09-II-CO-07-002
of the Far Eastern and Siberian Branches of Russian Academy of Science.
\end{acknowledgments}

\end{document}